\documentstyle[preprint,eqsecnum,aps,prb]{revtex}

\begin{document}
\draft
\title{Localized Basis for Effective Lattice Hamiltonians:\\
Lattice Wannier Functions}
\author{K. M. Rabe and U. V. Waghmare}
\address{Yale University, Department of Applied Physics\\
P. O. Box 208284, New Haven, Connecticut, 06520-8284}
\maketitle
\begin{abstract}

A systematic method is presented for constructing effective
Hamiltonians for general phonon-related structural transitions.
The key feature is the application of group theoretical methods
to identify the subspace in
which the effective Hamiltonian acts and construct for it localized basis
vectors, which are the analogue of
electronic Wannier functions.
The results of the symmetry analysis for the perovskite, rocksalt, fluorite and
A15 structures and the forms of effective Hamiltonians for the ferroelectric
transition in $PbTiO_3$ and $BaTiO_3$, the oxygen-octahedron rotation
transition in $SrTiO_3$,
the Jahn-Teller instability in $La_{1-x}(Ca,Sr,Ba)_xMnO_3$ and the
antiferroelectric transition in $PbZrO_3$ are discussed.
For the oxygen-octahedron rotation transition in $SrTiO_3$,
this method provides an alternative to the rotational variable approach which
is well behaved throughout the Brillouin zone.
The parameters appearing in the
Wannier basis vectors and in the effective Hamiltonian,
given by the corresponding invariant energy expansion, can be obtained for
individual materials using first-principles density-functional-theory total
energy and linear response techniques, or any technique that can reliably
calculate force constants and distortion energies.
A practical approach to the determination of these parameters
is presented and
the application to ferroelectric $PbTiO_3$ discussed.
\end{abstract}


\narrowtext

\section{INTRODUCTION}
Interest in the construction of effective Hamiltonians to relate macroscopic
properties of materials to microscopic information has recently increased
dramatically.
Construction of an effective lattice Hamiltonian is greatly facilitated by the
use of a localized basis for the description of crystal structure distortions.
For example, structural phase transitions in perovskite and rocksalt structure
systems have been studied using model Hamiltonians with a
``local mode'' basis to describe
the unstable lattice modes relevant to the transition.
While in earlier work the microscopic model parameters had to be
determined empirically,
\cite{muller,pytte,lines}
recent advances in the accuracy and efficiency of first-principles
bandstructure and total energy methods have made it possible to calculate these
parameters directly
from first principles,
and encouraging agreement with experimentally observed transition properties
has been obtained.\cite{abi,pbti,bati,srti}
In addition, the study of electronic properties of systems which exhibit
lattice instabilities coupled to the electronic states is facilitated by the
description of the lattice in a localized basis, putting it on an equal footing
with the usual tight-binding description of the electronic states.\cite{millis}
However, the approximate ``local mode'' basis is expected to be inadequate for
high-accuracy first-principles investigations.
Furthermore, explicit construction of a localized basis set and optimal
effective Hamiltonian has proved to be especially problematic for complicated
crystal structures and structural transitions involving modes at the zone
boundary or interior of the zone.

In this paper, we present a systematic procedure for the construction of an
effective lattice Hamiltonian in a localized basis set, applicable to arbitrary
systems and structures and specifically designed for practical use in
first-principles studies.
This approach is suitable for the investigation of a general phonon-related
structural transition,
\cite{mullers} involving a small change in structure which can be generated by
freezing phonon displacements into a high-symmetry
prototype structure.
We implement this procedure for structural transitions in the perovskite,
rocksalt, fluorite and A15 structures.
Group theoretical methods, which since Landau
\cite{landau} have been central to the phenomenological study of structural
phase transitions, are essential to our development of a procedure which is
systematic and easy to implement for arbitrary structures and transitions.
The resulting localized basis vectors are the lattice analogue
\cite{muller,pytte,kohn,eijn} of electronic Wannier
functions.\cite{kohna,descl}
These Wannier basis vectors span an invariant subspace containing the
low-frequency normal modes relevant to the transition.
This subspace is energetically decoupled at quadratic order from the higher
frequency eigenmodes, permitting the construction of an effective Hamiltonian
for the transition including only lattice distortions in the subspace.
The model parameters for a particular material can be obtained
using any method that can reliably calculate force constants and distortion
energies.
For the force constant calculation, the most efficient high-accuracy  method is
density-functional-theory (DFT) linear response.\cite{bgt,gonze,yale,krakauer}
As we will see, once the force-constant matrices have been calculated,  an
effective Hamiltonian can be constructed with very little additional effort.
The final result is a model system which reproduces the finite-temperature
transition behavior of the original lattice
Hamiltonian, while the simpler form and reduction in the number of degrees of
freedom per unit cell make it suitable for study by methods such as mean field
theory and Monte Carlo simulation.

In the next section, we present the derivation of the effective Hamiltonian in
a symmetrized localized (Wannier) basis. In Section III, we discuss in detail
the construction of the Wannier basis and resulting effective Hamiltonian for
the simple
example of the diatomic chain. In Section IV, this construction is
generalized to arbitrary structures and applied to the structural transitions
in perovskite compounds. In addition, the results of the construction for the
rocksalt, fluorite and A15 structures are briefly described.  Section V is a
discussion of the practical implementation of this approach in a
first-principles framework, illustrated by application to the ferroelectric
transition in $PbTiO_3$.
Finally, Section VI summarizes and concludes the paper.

\section{Formal derivation of the effective Hamiltonian}

Consider a crystal with $n_{at}$ atoms per unit cell and $N$ unit cells with
Born-von Karman boundary conditions.
Each possible configuration of the crystal is fully described by
specifying the generalized displacement of each ion (i.e. the displacement
multiplied by the square root of the ion mass) relative to a fixed
high-symmetry reference structure, which should be chosen so that
the physically important configurations are obtained by small distortions. In
general, this will be the same as the structure of the high-symmetry phase.
The set of all ionic displacement patterns forms a $3n_{at}N$-dimensional
vector space, called the ``ionic-displacement space."
In the standard basis, in which each basis vector is the unit displacement of a
single ion in a single cartesian direction,
the coordinates of a particular configuration are the
generalized displacements
$\{u_{j\alpha},~j=1,n_{at}N;~\alpha=x,y,z\}$.
Another basis for the ionic-displacement space is obtained from the
normal modes $\{e_{\vec k \lambda};~\vec k \in {\rm Brillouin~
zone},~\lambda=1,3n_{at}\}$.
In this basis, the coordinates $\{c_{\vec k \lambda}\}$ are complex numbers,
with
the understanding that the ionic-displacement pattern is obtained
by taking the real part of the vector $\sum_{\vec k, \lambda}c_{\vec k
\lambda}e_{\vec k \lambda}$.
The $\{e_{\vec k \lambda}\}$ transform according to irreducible
representations of the space group of the high-symmetry
reference structure ${\cal G}_0$. For example, the action of translation
through
a lattice vector is ${\cal O}_{\vec R}\;e_{\vec k \lambda}=exp(i\vec k \cdot
\vec R)e_{\vec k \lambda}$.

For the construction of an effective Hamiltonian, we perform a
decomposition of
the ionic-displacement space into invariant subspaces, called ``band
subspaces,'' spanned by one or more entire branches of the normal modes and
closed
under ${\cal G}_0$\cite{descl}
Each band subspace in this decomposition should be minimal in the
sense that it cannot be written as the direct sum of two band subspaces.
The number of band subspaces is $n_s$ and they are indexed by
$\Lambda$, with the number of branches in each subspace $n_\Lambda$, indexed by
$l$ (note that $\sum_{\Lambda=1,n_s}n_\Lambda=3n_{at}$).
$n_\Lambda$ will also be referred to as the multiplicity of the band subspace.

For each band subspace $\Lambda$, we seek a
localized basis set with definite
symmetry properties with respect to ${\cal G}_0$.
These properties are established by choosing a Wyckoff symbol for ${\cal G}_0$,
of multiplicity $n_w$, and an $l_s$-dimensional irreducible representation
(irrep) of the
corresponding site symmetry group.\cite{descl,hatch,stokes}
The choice of Wyckoff symbol and site symmetry group irrep is constrained by
the requirement that the representation of ${\cal G}_0$ defined by the
localized basis have the same decomposition into  irreps of ${\cal G}_0$ as the
particular band subspace $\Lambda$ for which a basis is being sought.
Specifically, a set of basis vectors $\{w_{\Lambda i j \vec R_n};~j=1,l_s\}$ is
associated with a particular site in the crystal
(the ith Wyckoff position added to the lattice vector $\vec R_n$) and is a set
of partner functions of the chosen irrep of the site symmetry group.
These basis vectors are localized, with ionic displacements
decaying rapidly with distance from the associated site.
The basis vectors associated with the other sites of the chosen type are
generated by acting on the given basis vectors with symmetry operations from
${\cal G}_0$. For example, subsets of localized basis vectors are
related to each other by translation through a lattice vector
$w_{\Lambda i j \vec R_n+\vec R}={\cal O}_{\vec R}\;w_{\Lambda i j \vec R_n}$.
Clearly, the product of the number of
Wyckoff positions $n_w$ and the dimension of the site symmetry group irrep
$l_s$ must equal the multiplicity of the band subspace $n_\Lambda$, and we
replace the former two indices by a combined index $l=1,n_\Lambda$.
The decomposition of the localized basis representation of ${\cal G}_0$ into
irreps of ${\cal G}_0$ can be computed from
formulas such as Equation 14 in Ref. \onlinecite{hatch} or by computing, for
various $\vec k$, the character table of ${\cal G}_{\vec k}$ in the subspace of
Bloch sums.
Usually, the appropriate choice of Wyckoff symbol and site symmetry group irrep
can be determined just from the space group irrep labels of the eigenmodes at
the high-symmetry points.

The basis vectors from each band subspace in the decomposition can be combined
to form a basis for the full ionic displacement space. In this basis, the
coordinates of a particular configuration are the real
numbers
$\{\xi_{\Lambda l \vec R_i};\Lambda=1,n_s;~l=1,n_\Lambda;~i=1,N\}$, which
determine the ionic-displacement pattern as the vector $\sum_{\Lambda l
i}\xi_{\Lambda l \vec R_i}w_{\Lambda l \vec R_i}$.
These localized basis vectors are the lattice analogue
\cite{muller,pytte,kohn,eijn} of electronic
Wannier functions.\cite{kohna,descl}
The symmetry properties of the Wannier basis vectors lead to a convenient form
for the Taylor expansion of the lattice energy in the corresponding
coordinates, allowing easy identification of independent parameters.
Furthermore, the expansion has the important feature that it contains no
quadratic cross terms between different band subspaces.
Extra considerations apply when the various branches of normal modes include
some with polar character and/or are not well separated in energy; these will
be discussed in Section V.

With this basis set, the derivation of an effective Hamiltonian proceeds as
follows.
The classical partition function for the lattice is:
\begin{equation}
Z \propto \int\{d u_{j\alpha}\} exp(-\beta {\cal H}_{lat}(\{u_{j\alpha}\}))
\label{part}
\end{equation}
where $j$ runs over the $n_{at}N$ atoms, the electron degrees of freedom have
been eliminated
through the Born-Oppenheimer approximation, and
the generalized ionic displacements $\{u_{j\alpha}\}$
are measured relative to a
high symmetry reference structure, as described above.
The momentum integral leads to a trivial $u$-independent factor and is not
explicitly written here.

It is straightforward to change variables in ${\cal H}_{lat}$ from standard
basis coordinates $\{u_{j\alpha}\}$ to Wannier basis coordinates
$\{\xi_{\Lambda l \vec R_i}\}$ and Taylor expand ${\cal H}_{lat}$ around the
reference structure.
By construction, the quadratic part ${\cal H}_{lat}^{(2)}$ decouples into
a sum over the band subspaces:
$${\cal H}_{lat}^{(2)}(\{\xi_{\Lambda l \vec R_i}\})
=\sum_{\Lambda=1,n_s}{\cal H}^{(2)}_\Lambda(\{\xi_{\Lambda l \vec R_i}\})$$
where
$${\cal H}^{(2)}_\Lambda(\{\xi_{\Lambda l \vec R_i}\})=
\sum_{l,l'=1,n_\Lambda}\sum_{i,j=1,N}a^{(\Lambda)}_{l i l' j}
\xi_{\Lambda l \vec R_i}\xi_{\Lambda l' \vec R_j}.$$

The next step is to choose a band subspace or direct sum of band subspaces as
the symmetry-invariant subspace in which the effective Hamiltonian will act.
This subspace should be as small as possible, yet should include all modes
which have higher-order terms in the expansion of $H_{lat}$ which are important
in describing low-energy configurations, and which
therefore cannot to a good approximation be dropped from the expansion in the
evaluation of the partition function near the transition.
For the calculation of thermodynamic behavior at temperatures in the vicinity
of the structural transition, it is essential to include higher order terms for
the branches containing the unstable modes associated with the transition.
Strain degrees of freedom may not have important anharmonicity in themselves,
but still have physically important higher-order coupling to the unstable
modes, and thus should be considered for inclusion in the subspace.
Here and in the rest of this section, we write the expressions for the case
that the effective Hamiltonian acts in a single band subspace $\Lambda_0$,
since generalization to the case of the direct sum
$\Lambda_1\oplus\Lambda_2\oplus...\oplus\Lambda_s$ is straightforward.
The correct choice of subspace permits the following approximation to
$H_{lat}$:
\begin{equation}
{\cal H}_{lat}\approx {\cal H}_{eff}(\{\xi_{\Lambda_0 l \vec R_i}\})
+\sum_{\Lambda \neq \Lambda_0}{\cal H}^{(2)}_\Lambda(\{\xi_{\Lambda l \vec
R_i}\})
\label{decoup}
\end{equation}
Within this approximation, anharmonic terms appear {\it only} in ${\cal
H}_{eff}$ and, by construction, there are {\it no} cross terms between
$\Lambda_0$ and $\Lambda \neq \Lambda_0$.
This has the important consequence that the integration over the  degrees of
freedom not included in the effective Hamiltonian subspace can be performed
analytically.
Specifically, the partition function integral factorizes into an integral over
$\Lambda_0$ and another over $\Lambda \neq \Lambda_0$.
The integral over $\Lambda \neq \Lambda_0$ is an elementary Gaussian integral
which leads to a temperature-dependent factor unimportant for the structural
transition. The remaining integral has the form of a partition function for a
system of $n_{\Lambda_0}$-component spins on a lattice $\{\vec R_i\}$, which
recovers the original partition function $Z$:
\begin{equation}
\int \{d\xi_{\Lambda_0 l \vec R_i}\}exp(-\beta {\cal H}_{eff}(\{\xi_{\Lambda_0
l \vec R_i}\}))\propto Z.
\label{second}
\end{equation}

An explicit expression for ${\cal H}_{eff}$ is obtained by making
a Taylor expansion in symmetry invariant combinations of the $\{\xi_{\Lambda_0
l \vec R_i}\}$. This expansion is greatly facilitated by the symmetry
properties of the Wannier basis.  An expression
with a finite number of undetermined parameters is obtained by setting
expansion coefficients beyond a certain range or order to zero or writing them
in a parametrized form. The resulting spin-like model is suitable for
statistical mechanical analysis via mean field theory, Monte Carlo simulation,
or other appropriate method. The values of the expansion coefficients are
different for different materials and are to be obtained from first principles
calculations, as will be discussed in Section V.

\section{Effective Hamiltonian construction: the diatomic chain}

In this section, the explicit construction of the effective Hamiltonian in a
localized Wannier basis will be presented for the simple example of the
diatomic chain.\cite{kohn}
This calculation demonstrates most, though not quite all, of the important
principles, including the determination of the symmetry properties of the
Wannier basis vectors.
The remaining generalizations for application to arbitrary structures will
be made in Section IV.

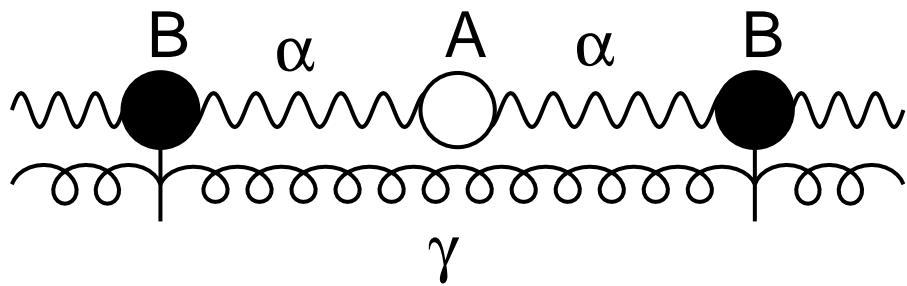
\begin{figure}
\caption{The harmonic potential energy of the diatomic chain model is shown by
two types of springs. One type, with spring constant $\alpha$, connects nearest
neighbor A ions (open circles) and B ions (solid circles). The other, with
spring constant $\gamma$,
connects nearest neighbor B ions. \label{abchain_fig}}
\end{figure}

The high-symmetry reference structure of the
``ionic'' diatomic chain model we consider has lattice constant $a=1$ and ions
A and B at positions
$n$ and $n+{1\over 2}$ ($n=1,...,~N$), respectively.
The potential energy in our model is given to quadratic order in
$\{u_{nA},u_{nB}\}$, the displacements from the reference structure, by
\begin{equation}
\sum_n[{\alpha\over 2}(u_{nB}-u_{nA})^2
+{\alpha\over 2}(u_{n-1,B}-u_{nA})^2
+{\gamma\over 2}(u_{n+1,B}-u_{nB})^2]
\label{chain}
\end{equation}
as shown in Fig. \ref{abchain_fig}.
In the rest of this section, we set the masses of the ions $m_A$ = $m_B$ = 1
and the ratio ${\gamma\over\alpha}=0.25$.

Analytical diagonalization of the dynamical matrix at each $k$ in the first
Brillouin zone yields two branches
($\lambda=1,2$) with eigenvalues $\epsilon_\Lambda(k)=\omega^2_\Lambda(k)$,
shown in Fig. \ref{abmodes_fig}.
The normal modes are labelled by the irreps of the space group
$\{\{U\vert n\hat x\},~U=E,I;~n=1,...,N\}$. These are $\Gamma_g$ and $\Gamma_u$
at $k=0$, $X_g$ and $X_u$ at $k=\pi$, and $k$ otherwise.
It can be seen that the ionic displacement space of the diatomic chain
decomposes into two band subspaces $\Lambda=1,2$, each spanned by the
nondegenerate eigenmode branch of corresponding $\lambda$.
Since the number of branches in each subspace is one, we drop the index $l$.

As we have seen in Section II, construction of an effective Hamiltonian
involves the choice of an invariant subspace. This subspace should be a band
subspace (or direct sum of band subspaces) which includes all modes relevant to
the particular transition being modelled.
In the diatomic chain example, we imagine that the relevant mode is $X_u$.
The band subspace $\Lambda=1$ has the symmetry label $X_u$, while the band
subspace $\Lambda=2$ has the symmetry label $X_g$. Therefore, the subspace in
which the effective Hamiltonian will act is $\Lambda=1$. If instead the
relevant mode were $X_g$, the corresponding effective Hamiltonian subspace
would be $\Lambda=2$.

For the chosen effective Hamiltonian subspace $\Lambda=1$ or $\Lambda=2$, we
seek a localized basis set labelled by a unit-multiplicity Wyckoff symbol and a
one-dimensional irrep of the corresponding site symmetry group, since both band
subspaces are of
unit multiplicity.
The space group of the diatomic chain has two Wyckoff symbols of unit
multiplicity: $1a$, (at x=0), and $1b$ (at x=1/2). For both $1a$
and $1b$, the site symmetry group is $\overline 1$, with two one-dimensional
irreps $A_g$ (even) and $A_u$ (odd).
The space group representations defined by each of the four distinct labels
($1a$,$A_g$), ($1a$,$A_u$), ($1b$,$A_g$) and ($1b$,$A_u$) decompose at the
high-symmetry k-points $k=0$ and $k=\pi$ into the space group irreps
($\Gamma_+$,$X_+$),
 ($\Gamma_-$,$X_-$),($\Gamma_+$,$X_-$), and ($\Gamma_-$,$X_+$), respectively.
Since the space group irrep labels of the localized basis set must match those
of the band subspace throughout the Brillouin zone, we find that for both band
subspaces the label of the Wannier basis is uniquely determined.
For $\Lambda=1$, it is ($1a$,$A_u$), and for $\Lambda=2$, it is ($1b$,$A_u$).

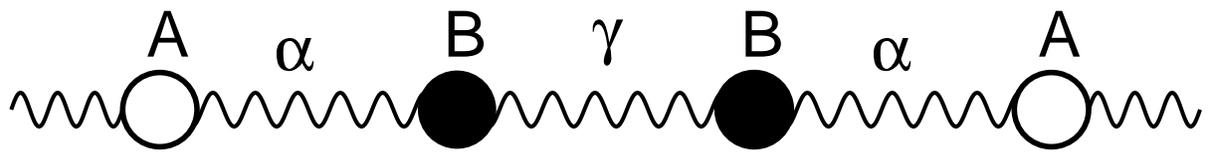
\begin{figure}
\caption{The harmonic potential energy of the triatomic chain model is shown by
two types of springs. One type, with spring constant $\alpha$, connects nearest
neighbor A ions (open circles) and B ions (solid circles). The other, with
spring constant $\gamma$,
connects nearest neighbor B ions. \label{triatomic_fig}}
\end{figure}

Since individual ionic displacements are a localized basis with the required
symmetry properties for the full ionic displacement space, it might seem that
the Wannier basis vectors would always be obtained from occupied Wyckoff
positions combined with vector representations of the site symmetry group, as
in the diatomic chain example.
In fact, this is not the case.
The simplest example in which this can be seen is the ``molecular'' diatomic
chain.\cite{kohn}
For the present discussion, we consider the closely related
triatomic chain model in Fig. \ref{triatomic_fig}. The high-symmetry reference
structure has lattice constant $a=1$ and ions A at positions $n$ and ions B at
positions $n\pm{1\over 3}$  ($n=1,...,~N$).
The space group is the same as that of the ``ionic'' diatomic chain discussed
above.
The space group irreps for the normal modes at the high-symmetry k-points are
$\Gamma_g$,
$\Gamma_u$, and $\Gamma_u$ at $k=0$ and $X_g$, $X_u$ and $X_u$ at $k=\pi$.
All space group irreps at the two points are compatible, so depending on the
details of the potential, four types of branches defining unit-multiplicity
band subspaces can appear.
The Wannier basis of the band subspace ($\Gamma_u$, $X_u$) is labelled by
($1a$,$A_u$), so that the basis vectors transform like ionic displacements at
occupied Wyckoff positions.
However, the Wannier basis of the band subspace ($\Gamma_g$,~$X_g$) is labelled
by ($1a$, $A_g$) and the Wannier bases of the two ``mixed-label'' band
subspaces ($\Gamma_u$,~$X_g$) and ($\Gamma_g$,~$X_u$) are labelled by the
unoccupied Wyckoff position $1b$, midway between the two B atoms, and the site
symmetry group irreps $A_u$ and $A_g$, respectively. For the triatomic chain,
with an appropriate choice of $\alpha$ and $\gamma$, it may happen that there
will be two band subspaces of the ($\Gamma_u$,~$X_u$) type.
Both will be described by Wannier basis vectors with labels ($1a$,$A_u$), but
the Wannier basis vectors for the two band subspaces will be orthogonal to each
other.

For the complete construction of an effective Hamiltonian, it is not enough to
know the symmetry properties of the basis vectors; rather, an explicit
expression for the Wannier basis vectors in terms of ionic
displacements is needed.
The basis vectors
are determined by the potential energy and resulting set of normal modes which
define the band subspace.
If the normal modes are known at all $k$, the basis vectors can be obtained
exactly.
For each $\Lambda$, the expression for $w_{\Lambda n}$, the Wannier basis
vector centered in unit cell $n$, in terms of $\{e_{k \Lambda}\}$ is
\begin{equation}
w_{\Lambda n}={1\over N}\sum_{-\pi< k \leq \pi}exp(i\phi_\Lambda (k))e_{k
\Lambda}exp(-ikn).
\label{ABwann}
\end{equation}
An inverse transform can be performed to express the $\{e_{k \Lambda}\}$ in
terms of the Wannier basis vectors:
\begin{equation}
e_{k \Lambda}=exp(-i\phi_\Lambda (k))\sum_n exp(ikn)w_{\Lambda n}.
\label{ABwanninv}
\end{equation}
In evaluating Equation \ref{ABwann}, the phase function $\phi_\Lambda (k)$ can
be freely chosen to maximize the localized character of $w_{\Lambda n}$, with
the only constraint being the requirement that the resulting Wannier basis
vectors have the desired symmetry properties.
For the diatomic chain, the normal modes are easily obtained at all $k$.
We follow the convention that the real part of the 2-component eigenvector
$(\Delta_{kA}^{(\lambda)},\Delta_{kB}^{(\lambda)})$ of the 2x2 dynamical matrix
at each $k$ gives the displacements of the A and B ions in the unit cell $n=0$.
The displacements of the A and B ions in the unit cell $n$ then are
$Re(\Delta_{kA}^{(\lambda)}exp(ikn))$ and
$Re(\Delta_{kB}^{(\lambda)}exp(ikn))$.
$\Delta_{kA}^{(\lambda)}$ and $\Delta_{kB}^{(\lambda)}$ can be chosen to
satisfy $\vert \Delta_{kA}^{(\lambda)} \vert^2+\vert \Delta_{kB}^{(\lambda)}
\vert^2=1$,
$\Delta_{kA}^{(\lambda)*}=\Delta_{-kA}^{(\lambda)}$ and
$\Delta_{kB}^{(\lambda)*}=\Delta_{-kB}^{(\lambda)}$ and to be smooth functions
of $k$, ensuring that the corresponding normal mode $e_{k \lambda}$ smoothly
depends on $k$.
For $\Lambda=1$, we choose $(\Delta_{kA},\Delta_{kB})$ such that
$\Delta_{kA}$ is real and positive for all $k$, and $\phi_\Lambda (k)=0$.
The resulting Wannier basis vector $w_{1n}$ is shown in Fig. \ref{aodd_fig}.

Except in special cases such as the diatomic chain, knowledge of the normal
modes at all $k$ is not available and the expression in Equation \ref{ABwann}
cannot
be used to construct the Wannier basis vectors.
However, a good approximation can be obtained from knowledge of the
eigenvectors at a subset of k-points through a fitting procedure
which will prove to be practical for use with first-principles calculations.
For a particular choice of Wyckoff symbol and site symmetry irrep, a
parametrization of the displacement pattern of the Wannier basis vectors is
obtained from symmetry considerations.
For example, for the Wannier basis ($1a$, $A_u$) of the $\Lambda=1$ subspace of
the diatomic chain, the displacement of the central ion
is given by one real parameter and each pair of ions equidistant from the
center requires an additional real parameter to specify the equal displacements
(see Fig. \ref{aodd_fig}).
The parameters of the Wannier basis vector, starting
with the innermost coordination shells, are then fit to reproduce a set of
known normal modes in the band subspace.
Since the Wannier basis vector is localized, it should be a good approximation
to set the displacement parameters to zero beyond a finite number of shells.
For example, in the diatomic chain, the simplest nontrivial truncation is at
the first shell. For $\Lambda=1$, the fit to the eigenvector at $k=0$ yields
the approximate Wannier basis vectors $w^{(0)}_{1n}$  with displacements
${1\over\sqrt 2}$ for the central A ion and ${1\over2\sqrt 2}$ for the first
shell of B ions.
The frequencies corresponding to the approximate normal modes $\sum_n
exp(ikn)w^{(0)}_{1n}$, computed from Eqn. \ref{chain} and shown in Fig.
\ref{abwapp_fig}, agree very well with the true frquencies if the
nonorthonormality of the basis vectors is taken into account.
With the inclusion of just one more shell of A ions in the approximate Wannier
basis vector to fit the normalized eigenmode at $k=\pi$ shown in Fig.
\ref{abmodes_fig},
it is nearly indistinguishable from the exact Wannier basis vector on the scale
of Fig. \ref{aodd_fig} and
unit normalization to within a few percent is achieved throughout the Brillouin
zone.
This construction also illustrates the point that the choice of phases for the
eigenmodes is important in producing a well-localized Wannier basis vector. In
this case, if the normalized eigenmode at $k=\pi$ were multiplied by $-1$, the
approximate Wannier basis vector would have larger displacements on the first A
coordination shell  than on the central A ion.

Using the symmetry properties of the Wannier basis vectors, the form of the
effective Hamiltonian can be written as an invariant expansion of the energy in
the Wannier basis coordinates of the subspace.
We now consider the diatomic chain with effective Hamiltonian subspace
$\Lambda=1$.
Since these coordinates have the same transformation properties as individual
ionic displacements of the A ions, the effective Hamiltonian parameters can be
put into direct
correspondence with the familiar force constants of the monoatomic chain or
with tight-binding parameters for a system with one $p$ orbital per unit cell.
Truncating the expansion at fourth order in onsite terms and at quadratic order
and second neighbor in intersite terms yields:
\begin{equation}
{\cal H}_{eff}(\{\xi_{\Lambda n}\})=\sum_n [a_0 \xi_{\Lambda n}^2+b_0
\xi_{\Lambda n}^4] + \sum_{n, n' s.t. \vert n-n' \vert \leq 2}a_{\vert n-n'
\vert} \xi_{\Lambda n} \xi_{\Lambda n'}
\label{ABmod}
\end{equation}
\noindent
The small number of expansion coefficients in this effective Hamiltonian can be
determined from total energy data for a particular choice of $\alpha$ and
$\gamma$.
Specifically, once the subspace Wannier basis vectors $\{w_{\Lambda n}\}$ have
been determined, the values of the coordinates $\{\xi_{\Lambda n}\}$ specify a
unique ionic
configuration.
${\cal H}_{eff}(\{\xi_{\Lambda n}\})$ is then taken as the potential energy of
this configuration.
Calculation of the coefficients $a_n$ in the invariant expansion proceeds by
fitting the sum of quadratic terms in Equation \ref{ABmod} for coordinates
$\{\xi_{\Lambda n}=Re(\xi_k\;exp(ikn))\}$, at various $k$, to the quadratic
energy of the corresponding ionic configurations, easily obtained from the
dynamical matrix derived from Equation \ref{chain}.
For example, when ${\gamma\over\alpha}=0.25$ the quadratic energy at $k=0$,
$k={\pi\over 2}$ and $k=\pi$ can be fit by $a_0=(13-\sqrt{33})\alpha/16$,
$a_1=-\alpha/4$ and $a_2=(-5+\sqrt{33})\alpha/32$.
Comparison of the quadratic part of the effective Hamiltonian, truncated at
$a_2$, with the true quadratic energy at other $k$ is used to check the
accuracy of the truncation of the invariant expansion, which in this case is
within a few percent.
Anharmonic parameters such as $b_0$ are obtained from calculation and fitting
of full potential energies for finite distortions.

In the study of structural transitions, focus on a simple description of a
relevant mode at a particular k-point has occasionally led to a choice of
localized basis vectors inconsistent with this symmetry criterion.
As will be discussed in Section IV, a common example is the use of a rotational
variable to describe the $R_{25}$ instability in perovskites.
The negative consequences of inconsistent symmetry can be simply illustrated in
the diatomic chain.
The displacement pattern of the $X_g$  mode (Fig. \ref{abmodes_fig}) can be
obtained with localized basis vectors not only of the type ($1b$,$A_u$) but
also ($1a$,$A_g$), the latter being
inconsistent with the symmetry criterion.
As shown above, the choice of ($1b$,$A_u$) is the label of the Wannier basis
for the $\Lambda_2$ band subspace.
However, localized basis vectors of the type ($1a$,$A_g$) lead to zero ionic
displacements for $k=0$ distortions, showing that this ``Wannier basis set''
is in fact linearly dependent.

The problems encountered when using an inconsistent symmetry  choice in the
construction of the effective Hamiltonian can be understood using Equations
\ref{ABwann} and \ref{ABwanninv}.
Wannier-like basis vectors inconsistent with our symmetry criterion can be
constructed only if the requirement that the inverse transforms (Equation
\ref{ABwanninv}) reproduce {\it normalized} normal modes is lifted and zero
normalization allowed.
In Equation \ref{ABwann}, this corresponds to multiplying the phase factor
$exp(i\phi_\Lambda (k))$ by a real function $f_\Lambda(k)$.
For example, Wannier-like functions of the type ($1a$,$A_g$) for the upper
branch ($\Lambda=2$) can then be obtained with $\phi_\Lambda (k)={\pi\over 2}$
and $f_\Lambda(k)={k\over \pi}$.
The inverse transforms reproduce the eigenmodes $i{k\over \pi}e_{k \Lambda}$,
not normalized to one except at $k=\pi$.
The energy written in terms of the Fourier transform of the Wannier basis
coordinates $\xi_\Lambda (k) = \sum_n exp(ikn)\xi_{\Lambda n}$ is
$$\sum_k {k^2\over \pi^2}\epsilon_\Lambda(k)\vert \xi_\Lambda(k)\vert^2$$
Effective Hamiltonian parameters can be fit as above to reproduce the
dispersion ${k^2\over \pi^2}\epsilon_\Lambda(k)$.
In particular, at $k=0$, the parameters must be fit to zero energy.
To use material-specific information at $k=0$, it is necessary to expand the
model dispersion to leading order in $k^2$ and to fit the coefficient of
${k^2\over \pi^2}$ to $\epsilon_\Lambda(k=0)$.
The approximate Wannier basis vector construction can also be carried out as
above except at $k=0$, where it is necessary to expand the approximate
eigenmode to leading order in $k$ and to fit the coefficients to
$e_{k=0,\Lambda}$.
With $f_\Lambda(k=0)=0$, required to eliminate the inconsistent condition on
the phases at $k=0$, the resulting localized basis is linearly dependent
and cannot reproduce the normal mode at $k=0$.
This amounts to eliminating the point $k=0$ from the ionic displacement space.
In the thermodynamic limit, the behavior of the system should not be affected,
since
in the partition function configuration integral, this is a set of measure
zero. However, in a finite-size simulation, the integral is replaced by a sum,
and the incorrect description of the system arising from the linear dependence
leads to an ${\cal O}({1\over N})$ error.
If transformation properties consistent with our symmetry criterion are chosen
and the normalization requirement imposed, these finite-size simulation errors
and the unnecessary complications in the first-principles fitting procedure are
avoided.

\section{Effective Hamiltonians for cubic systems}

In this section, we generalize the construction of the effective
Hamiltonian to arbitrary structures and apply it to phonon-related
structural transitions in the perovskite, rocksalt, fluorite and A15
structures.
We will use as our primary illustration the example of the perovskite
structure.
Among the many compounds in this structure, a rich variety of phonon-related
structural transitions are found.\cite{mullers} These can be analyzed in a
unified systematic framework using the present approach.

The cubic perovskite structure, which will be taken as the high-symmetry
reference structure, has five atoms per unit cell and space group $Pm3m$.
The first step in the effective Hamiltonian construction is a symmetry analysis
to identify the band subspaces (i.e. minimal invariant subspaces spanned by one
or more entire branches of normal modes).
Once the space group irrep labels for the ionic-displacement space have been
determined, a complete list of the band subspaces and the
corresponding Wannier basis labels can be obtained by the following general
procedure, valid for arbitrary structures.
First, all possible symmetry types of band subspaces are constructed
using the compatibility relations between the irrep labels at the
high-symmetry $\vec k$-points.
The corresponding Wannier basis labels are found by
considering in turn eachWyckoff symbol with each irreducible representation of
the corresponding site symmetry group such that the
product of the number of the Wyckoff positions with the dimension of the site
symmetry group irrep is less than or equal to the maximum
multiplicity of the band subspaces.
A particular combination will label a Wannier basis of a band subspace if the
space group irrep labels produced at the high-symmetry $\vec k$-points match
those of the band subspace.\cite{isotropy}
Application of this procedure to the perovskite structure yields
four band subspaces, all labelled by occupied Wyckoff positions with vector
representations, corresponding to atomic displacements. These are listed in
Table I.

As in the case of the diatomic chain, a parametrization of the displacement
pattern of the Wannier basis vectors can be obtained from symmetry
considerations.
For each band subspace, the procedure is (1) choose one of the Wyckoff
positions specified by the label; (2) identify the symmetric coordination
shells surrounding that site; and (3) compute the independent displacement
patterns of each shell that transform according to the given irreducible
representation of the site symmetry group.
For each of the four band subspaces of the perovskite structure, the
displacement patterns for the innermost coordination
shells are shown in Fig. \ref{pattern_fig}.
The relative amplitudes of each independent displacement pattern for each shell
are free parameters determined by the requirement that a particular branch or
branches of normal modes be reproduced, and thus will be different for
individual materials.

As we have seen in Section II, construction of an effective Hamiltonian
involves the choice of an invariant subspace as the direct sum of one or more
band subspaces. The effective Hamiltonian subspace should include all modes
relevant to the particular transition being modelled.
The smallest subspace satisfying this condition, and therefore the maximum
possible reduction in the degrees of freedom, can be identified by listing the
symmetry labels of the relevant modes and choosing a direct sum of the band
subspaces of the ionic-displacement space which produces the desired labels.
There are cases in which there is more than one minimal choice which satisfies
the conditions. If the choices differ in the other symmetry labels, the choice
is made for an individual material by determining the symmetry labels of the
lowest-energy modes at other high-symmetry $\vec k$-points. If all symmetry
labels are the same, then the Berry phase label \cite{zak,zakp} determines the
choice, since well-localized Wannier basis vectors are only obtained by
transforming eigenvector branches which are smooth in $\vec k$.
The general analysis of Berry phases for composite energy bands is quite
complicated.\cite{zaku}
For present purposes, we observe that if only one of the possible choices is
produced by an occupied Wyckoff position combined with a vector representation,
that choice will smoothly reproduce the physical eigenvector branch.
Examples of systems where more than one choice is of this kind can be
constructed, the simplest being a crystal of space group $F222$ (\# 22)
\cite{bacry}
with both the 2a and 2b Wyckoff positions occupied.
In such cases, it is advisable to include both branches in the effective
Hamiltonian subspace, with two Wannier basis vectors, one centered on each
Wyckoff position.
Only if the ``relevant phonon'' is heavily dominated by displacements of only
one of the two atom types should construction of a subspace with only one
Wannier basis vector (centered on the corresponding Wyckoff position) be
considered.
In the crystal structures analyzed in this section, this difficulty does not
arise.

With the selection of the effective Hamiltonian subspace, the invariant
expansion of the lattice energy in Wannier basis coordinates proceeds
straightforwardly, facilitated by the symmetry properties of the Wannier basis
vectors.
Different structural transitions observed in perovskite structure compounds
require different treatment, starting at this point.
In the following, we will consider four common perovskite transitions
(i) the ferroelectric transition in $BaTiO_3$ and $PbTiO_3$, with an unstable
$\Gamma_{15}$ mode;
(ii) the oxygen-octahedron rotational transition in $SrTiO_3$, with an unstable
$R_{25}$ mode;
(iii) the Jahn-Teller instability in $La_{1-x}(Ca,Sr,Ba)_xMnO_3$, with an
unstable $R_{2}'$ mode; and
(iv) the antiferroelectric transition in $PbZrO_3$, involving several unstable
modes.

For the $\Gamma_{15}$ ferroelectric transition found in $BaTiO_3$ and
$PbTiO_3$, all four subspace choices $A$, $B$, $O_{x,x}$ and
$O_{x,y}$ are consistent with the symmetry
criterion.
The simplest form of the effective Hamiltonian is obtained
by choosing either the $3N$-dimensional subspace $A$, resulting
in the invariant energy expansion for $PbTiO_3$ of Ref. \onlinecite{pbti}, or
the $3N$-dimensional subspace $B$ for $BaTiO_3$ as in Ref. \onlinecite{bati}.
In these two cases, the choice of $A$ versus $B$ is determined by the symmetry
labels of the lowest zone-boundary modes.
In these systems, coupling to strain is known to be important.
\cite{pbti,cohen,kingsmith}
Homogeneous strain is easily included by introducing the six-component strain
variable $e_{\alpha \beta}$ ($\alpha, \beta=x,y,z$) and its lowest order
coupling to the ferroelectric mode coordinates.\cite{abi,pbti,bati}
To include inhomogeneous strain as well, the subspace can be enlarged by adding
a second band subspace containing $\Gamma_{15}$ to describe the acoustic modes.
The most convenient choice is the cation subspace not already used for the
ferroelectric mode.
In addition to the space group symmetries, the inhomogeneous strain Hamiltonian
has global translation and rotation invariances which can most easily be
satisfied by following the construction method of Keating.
\cite{keating}
If the expansion is truncated at three independent parameters, corresponding to
the three elastic constants, the strain Hamiltonian used in Ref.
\onlinecite{bati} is obtained. Additional expansion parameters could be
included to improve the description of the acoustic modes away from the zone
center.
With Wannier basis vectors that exactly reproduce the mode eigenvectors, there
is no quadratic coupling between the two subspaces $A$ and $B$. The lowest
order anharmonic coupling that does not vanish in the limit $\vec k \rightarrow
0$ is
the coupling term used in Ref. \onlinecite{bati}.

The $R_{25}$ oxygen-octahedron rotational transition occurs in $SrTiO_3$ at 105
K and in a number of other systems.\cite{lg}
This unstable mode is built up in Refs. \onlinecite{pytte}~and
\onlinecite{srti} from a Wannier-like ``rotational variable" transforming
according to
$T_{1g}$
\cite{burns} at each site of the simple cubic lattice. Symmetry analysis shows
that this basis set is linearly dependent, generating symmetry labels
$\Gamma'_{25}$, $\Delta'_{2}$ and $X_{2}$ which do not appear in the symmetry
decomposition of the ionic displacement space. While this difficulty can be
formally resolved by requiring the energy to be zero for configurations with
$\vec k$ in
$\Gamma-\Delta-X$, the vanishing of the normalization at lines in the Brillouin
zone leads to ${\cal O}({1\over N^2})$ errors in finite-size simulations, as
discussed in the previous section for the diatomic chain. Also, incorporation
into the effective Hamiltonian of first-principles results at these lines
requires extra manipulations. Finally, use of this basis in quantum-mechanical
or dynamical studies, in which the kinetic energy must be treated explicitly,
would complicate the analysis by introducing an artificially vanishing
effective mass in the neighborhood of these lines.

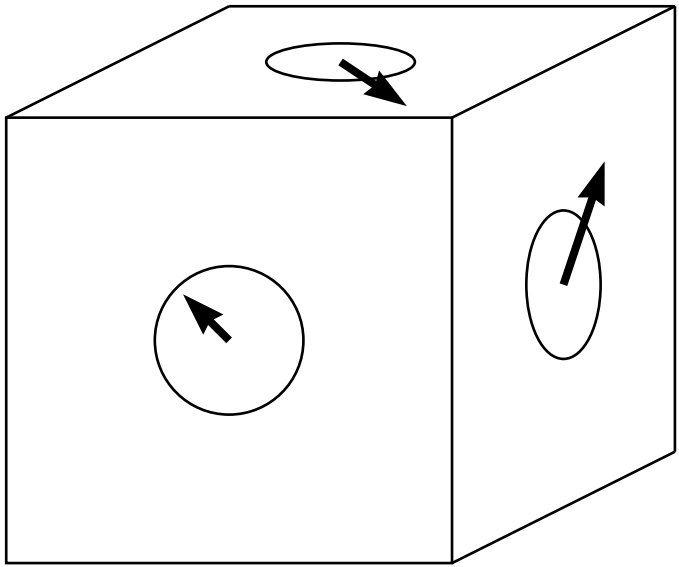
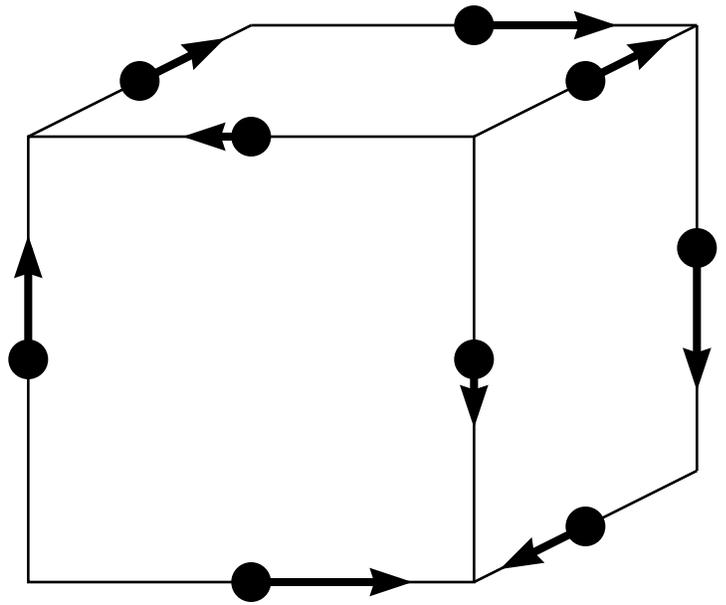
\begin{figure}
\caption{(a) The model system for the $R_{25}$ oxygen-octahedron rotational
transition in $SrTiO_3$ consists of two-component vectors on the faces of the
unit cell of a simple cubic lattice;
(b) the model system for the $R'_{2}$ Jahn-Teller instability in
$La_{1-x}(Ca,Sr,Ba)_xMnO_3$ consists of one-component vectors on the edges of
the unit cells of a simple cubic lattice. \label{hello_fig}}
\end{figure}

In the present approach, to describe the $R_{25}$ oxygen-octahedron rotational
transition in $SrTiO_3$, the band subspace is the $6N$ dimensional space
$O_{x,y}$. The resulting effective Hamiltonian describes a system of
two-component vectors on the faces of the unit cells of the simple cubic
lattice (Fig. \ref{hello_fig}(a)).
This form is slightly more complicated than the models usually studied in the
statistical mechanics literature, since
the ``spin'' sites do not form a Bravais lattice.
However, symmetry still ensures that the invariant energy expansion
is fairly simple. In fact, the quadratic part has the form of a submatrix of
the familiar force constant matrix for the perovskite structure.
Furthermore, this form presents no problems for numerical analysis of the model
through Monte Carlo simulation.
Even the fact that the band subspace is six dimensional can be turned to
advantage. The extra $\Gamma_{15}$ mode can be used to describe inhomogeneous
strain coupling or the ferroelectric mode, which is observed to be nearly
unstable.\cite{lga}

To describe the $R'_{2}$ Jahn-Teller instability in $La_{1-x}(Ca,Sr,Ba)_xMnO_3$
\cite{kanamori}
the invariant subspace for effective Hamiltonian construction must include the
$3N$ dimensional band subspace labelled $O_{x,x}$ in Table I.
The resulting effective Hamiltonian describes a system of one-component vectors
on the edges of the unit cells of a simple cubic lattice.
As in the case of $SrTiO_3$, the invariant energy expansion turns out to be
simple and quite tractable for analysis through Monte Carlo simulation. The
symmetry of the model can also be readily exploited in writing the form of the
coupling to electronic states.\cite{millis}

For antiferroelectric $PbZrO_3$, there are four relevant modes:
an $R_{25}$ oxygen-octahedron rotation, a $\Gamma_{15}$ polar mode, a $({1\over
2} {1\over 2} 0)({\pi \over a})\Sigma_3$ mode, and a $(110)({\pi \over a})M'_5$
mode.\cite{cochran} All these modes are included in the $6N$-dimensional band
subspace $O_{x,y}$ (Fig. \ref{hello_fig}(a)), so that the resulting effective
Hamiltonian is surprisingly simple for such a complex transition. To examine
the effects of coupling to inhomogeneous strain,
the effective Hamiltonian subspace would have to be extended to include an
additional $3N$-dimensional subspace ($A$ or $B$).

The analysis for structural transitions occurring in other structures proceeds
similarly.
The rocksalt-rhombohedral transition in $GeTe$ has been previously studied with
a first-principles model Hamiltonian approach.\cite{abi}
The high-symmetry reference structure is the rocksalt structure, with an $fcc$
Bravais lattice with conventional lattice constant $a_0$ and space group
$Fm3m$.
The symmetry analysis of the
band subspaces is given in Table II.
The soft mode of the transition to the rhombohedral structure is $\Gamma_{15}$,
so
both $Ge$ and $Te$ are possible choices for the effective Hamiltonian subspace.
As in the perovskite structure, the choice is determined by the symmetry label
of the lowest energy zone-boundary mode.
The effective Hamiltonian constructed with the $Te$ subspace is a system of
three-component spins on an $fcc$ lattice of lattice constant ${a_0}$. In Ref.
\onlinecite{abi}, coupling to homogeneous strain was included and good
agreement obtained with the experimental transition behavior.
The rocksalt structure is so simple that if inhomogeneous strain were
included in the model (by adding the $Ge$ subspace) the resulting ``effective
Hamiltonian'' subspace would be the full ionic displacement space, and the
effective Hamiltonian would be identical to ${\cal H}_{lat}$.

The cubic-tetragonal transition in $ZrO_2$ has inspired several
recent first-principles bandstructure and total-energy studies.\cite{zirconia}
In this system, the high-symmetry reference structure is the fluorite
structure, with an $fcc$ Bravais lattice with conventional lattice constant
$a_0$ and space group $Fm3m$.
The symmetry analysis of the
band subspaces is given in Table III.
The soft mode of the transition to the tetragonal structure is $X'_2$,
so the effective Hamiltonian subspace must be the $6N$-dimensional space $O$.
The resulting effective Hamiltonian is a system of three-component spins on a
simple cubic lattice of lattice constant ${a_0\over 2}$.
However, the symmetry for the invariant energy expansion is not
$O_h$, but $T_d$, imposed by the presence of the $fcc$ zirconium sublattice.

Structural transitions in A15 compounds such as $V_3Si$ have long been of
interest because of the interplay with ``high-temperature''
superconductivity in these systems.\cite{a15str} The A15 structure has two
formula units (8 atoms) per unit cell, with a simple cubic Bravais lattice and
a nonsymmorphic space group $Pm3n$.
The symmetry analysis of the
band subspaces is given in Table IV.
The observed low-temperature structure  is generated by the $\Gamma_{12}$ mode.
The effective Hamiltonian subspace is then the $6N$-dimensional space
$A_{\parallel}$.
The choice of $A_{\parallel}$ over $6d,{\vec u_\parallel}$, which also produces
the right symmetry label, is made on the basis of the Berry phase argument
discussed earlier in this section.
The resulting model is a system of one-component vectors on three
interpenetrating sets of chains running along the cartesian directions. The
$\Gamma_{15}$ mode in this subspace can be used to include the effects of
coupling to inhomogeneous strain.

\section{Practical aspects of first-principles implementation}

The information available from first-principles methods is (i) total energy of
supercells; and (ii) force constant matrices at various $\vec k$ calculated
using density functional theory linear response.\cite{bgt,gonze,yale,krakauer}
In this section, we present a practical approach for using this information to
construct the effective Hamiltonian for a particular system, illustrating the
procedure with the application to the ferroelectric transition in
$PbTiO_3$.\cite{pbti}

The analysis begins with the identification of the high-symmetry reference
structure and relevant phonon, and the
first-principles calculation of the force constant matrices (either by linear
response or by frozen phonon supercell calculations) at the high-symmetry $\vec
k$-points.
The first step is to identify the effective Hamiltonian subspace
$\Lambda_0$.
For some structural transitions,
$\Lambda_0$ will be a single band subspace uniquely specified by the relevant
phonon.
For others, the relevant phonon allows a choice of band subspaces.
In the example of the $PbTiO_3$ transition, the $\Gamma_{15}$ phonon allows a
choice of band subspaces $A$, $B$, $O_{x,x}$ and $O_{x,y}$.
The choice should be made so as to include the lowest energy eigenmodes at
other high-symmetry $\vec k$-points.
In $PbTiO_3$, the low energy modes $X'_5$, $M'_2$, $M'_5$, and $R_{15}$
determine the efffective Hamiltonian subspace to be the band subspace $A$ (see
Table \ref{tablea}).

The next step is the construction of the Wannier basis vectors for the subspace
$\Lambda_0$.
The strategy is to fit independent displacement parameters for the innermost
coordination shells to the normalized eigenvectors at only the high-symmetry
$\vec k$-points
(with adjustment of the eigenvector phases to obtain optimally rapid decay).
Since the Wannier basis vector ionic displacements generally decay rather
rapidly with distance from the central ion, this is sufficient to determine an
excellent approximation to the Wannier basis vector. In the example of the
diatomic chain, discussed in Section III,
it was sufficient to fit only the $k=0$ and $k=\pi$ modes.
The application to ionic insulating crystals, including $PbTiO_3$, requires
additional comment.
If $\Lambda_0$ contains a degenerate polar $k=0$ mode and there are polar $k=0$
modes not included in $\Lambda_0$,
(as, for example, in the perovskite structure
\cite{zhongv}), the electric field at $k\rightarrow0$ mixes the LO modes
differently from the TO modes, and a single Wannier basis vector cannot
faithfully reproduce both the transverse and longitudinal branches.
Therefore, the normal modes of the corresponding longitudinal branch should
{\it not} be used in constructing the approximate Wannier basis vector. The
correct approach is to use only the low energy (transverse) branches to
determine the approximate Wannier basis vector.
For $PbTiO_3$, we use displacements of the central, first and second neighbor
Pb ions and the nearest Ti and O shells, for a total of nine independent
parameters
(six of which are included in Fig. \ref{pattern_fig}) to reproduce the
normalized lowest-frequency normal modes $\Gamma_{15}$, $R_{15}$ and $M'_2$ and
the normalization of the $M'_5$ mode exactly. The amplitudes of the
displacements are found to decay rapidly, as expected. Moreover, the $X'_5$ and
$M'_5$ normal modes predicted from the parameters obtained from the fit are in
excellent agreement with the corresponding normalized
normal modes calculated from first principles.\cite{pbti}

The form of the effective Hamiltonian is obtained by Taylor expanding the
potential energy in symmetry invariant combinations of the $\xi_i$ and
approximating to obtain an expression with a relatively small number of
effective Hamiltonian parameters which can be determined from the available
information.
For the quadratic part of the effective Hamiltonian, this will mean that
intercell interactions of the most general form are included only up to a
finite range.
For $PbTiO_3$, we include general intercell interactions up to third neighbor
(eight independent parameters).
The long-range dipolar interactions required for the correct description of
polar modes can be included in a parametrized form, requiring only knowledge of
the Born effective charges $Z^*$ and the dielectric constant $\epsilon_\infty$,
which can be obtained from separate linear response calculations or from
energies of distortions with $\vec q$ close to zero.
Similarly, a small number of higher order terms must be selected so as to
ensure stability and a ground state configuration of the correct type.
In $PbTiO_3$, only onsite anharmonic terms are needed.
Lastly, in a cubic system such as $PbTiO_3$, inclusion of the coupling to
homogeneous strain introduces only a few additional parameters.

The calculation of the parameters in the effective Hamiltonian exploits the
fact that a given set of Wannier basis coordinates uniquely determines the
positions of the ions.
The quadratic parameters are obtained by fitting to the frequencies of a set of
modes in the effective Hamiltonian subspace computed from first-principles
force-constant matrices at the high-symmetry points $\vec k$-points and at
$\vec k$-points along selected symmetry lines.
In $PbTiO_3$, we compute $Z^*$ and $\epsilon_\infty$ from DFT perturbation
theory, and fit the eight short-range interaction parameters to the following
effective Hamiltonian eigenmodes: $\{\Gamma_{15},R_{15},M'_2,M'_5,X'_2,X'_5\}$
and the $\Lambda_3$ modes at ${\pi\over 2a_0}(111)$ and ${\pi\over 4a_0}(111)$.
The truncations in the Taylor expansion of the quadratic effective Hamiltonian
can be checked by computing force constant matrices at additional $\vec k$.
With linear response, this overdetermination of parameters is relatively easy,
while it can be quite impractical in supercell calculations.
In the same way, anharmonic terms can be obtained from fitting the higher-order
dependence of first-principles total energies on the amplitude of the Wannier
basis coordinates.
Lastly, the homogeneous strain terms are obtained by computing the dependence
of total energies of configurations on variations in the lattice parameters.

The approximate Wannier basis vector approach can be applied successfully even
if the branches of interest are not fully isolated in energy from other
branches, which are recognizable by distinct symmetry labels.
Even in this case, a simple model, with the effective Hamiltonian acting in the
$\Lambda_0$ subspace, should still give accurate results for finite-temperature
properties if the
$\Lambda\neq\Lambda_0$ branches cross only the higher-energy portions of the
$\Lambda_0$ branches, preferably at $\vec k$-points far from the $\vec k$ of
the relevant phonon.
In that case, the approximate Wannier basis vector is fit using only the
normalized eigenvectors at $\vec k$-points where the $\Lambda_0$ branches can
be distinguished from the $\Lambda\neq\Lambda_0$ branches by their symmetry
labels. This set of $\vec k$-points will typically be the same as the set of
high-symmetry $\vec k$-points prescribed in the general procedure for
approximate Wannier basis vector construction.
Similarly, at other $\vec k$-points, energies of the approximate eigenvectors
built up from these approximate Wannier basis vectors are computed from force
constant matrix results and used to determine effective Hamiltonian parameters.
Therefore, application of the approximate Wannier basis vector approach
essentially projects out the $\Lambda\neq\Lambda_0$
character at the low-symmetry $\vec k$-points, which amounts to ignoring the
crossing.

If a substantial amount of crossing occurs into the lower energy range, if all
the relevant modes are not contained in a single band subspace,
or if anharmonic coupling to modes outside the relevant band subspace
is important, it will be necessary to enlarge the effective Hamiltonian
subspace to include the additional branches by working with a direct sum
$\Lambda_1 \oplus  \Lambda_2 \oplus...\oplus\Lambda_s$.
For each band subspace, there is an independent Wannier basis vector to be
determined by fitting eigenvector information.
If the approximate Wannier basis vectors exactly reproduce all branches at all
calculated $\vec k$-points, then in the invariant energy expansion there are no
cross terms between coordinates from different band subspaces.
If this is not the case (for example, for the polar modes discussed above)
cross terms must be included in the invariant energy expansion and explicitly
determined from the force constant matrix information.

If the subspace containing the relevant modes contains no polar modes and is
well separated in energy from the higher-energy eigenmodes, there is an
alternative approach which bypasses the explicit construction of the Wannier
basis vectors altogether by applying the philosophy of tight-binding
parameterization of electronic energy bands.\cite{tbspirit} In that case, we
assume the existence of underlying Wannier basis vectors and fit directly to
the eigenvalues to obtain effective Hamiltonian parameters.
This approach was discussed in Section III for the diatomic chain example,
which satisfies the necessary conditions.
In applications to real materials, this ideally simple situation is not
expected to be very common, and therefore the approximate Wannier basis vector
construction is recommended for routine use.

\section{Conclusions}

A systematic procedure has been presented for the construction of localized
Wannier basis vectors and corresponding effective Hamiltonians,
applicable to arbitrary structures and phonon-related transitions.
The application to several different structures has been described.
For the perovskite structure, a new approach to the oxygen-octahedron
transition in $SrTiO_3$ is suggested, which avoids the subtle difficulties
arising from the use of a ``rotational variable."
Improvements can be systematically made by
including additional band subspaces in the effective Hamiltonian subspace, if
indicated by the presence of substantial
crossing of eigenmode branches or strong anharmonic coupling to inhomogeneous
strain or other modes.
The effective Hamiltonian parameters for a particular material can be obtained
using any method that can reliably calculate force constants and distortion
energies.
Once the force-constant matrices have been calculated,  the construction of an
effective Hamiltonian is straightforward and requires very little additional
computational effort.
An efficient method, such as DFT linear-response, expedites the necessary
overdetermination of effective Hamiltonian parameters.
Further applications of this approach could include the construction of
realistic quantum mechanical models for the study of quantum paraelectrics and
ferroelectrics,
\cite{quantum} as well as classical dynamical models for the study of
ferroelectric phenomena such as switching.
In addition,
these effective Hamiltonians can be used for realistic first-principles studies
of thermal expansion and other temperature dependent lattice properties,
\cite{sugino} as well as for structural transitions.
More generally, the construction of a Wannier basis may prove advantageous for
the treatment of localized perturbations.
\vskip 0.2in
\centerline{\bf Acknowledgments}

We thank R. B. Phillips, G. Moore, S. Ogut, A. J. Millis, A. Rappe and W. Zhong
for useful discussions. We thank D. Vanderbilt, H. T. Stokes, and D. M. Hatch
for useful discussions and for valuable comments on the manuscript.
This work was supported by ONR Grant N00014-91-J-1247. In addition, K. M. R.
is grateful for the hospitality of l'Ecole d'Et\'e de Physique Th\'eorique Les
Houches during part of this work, and for the support of the Clare Boothe Luce
Fund and the Alfred P. Sloan Foundation.

\narrowtext
\begin{table}
\caption{Band subspaces of the ionic displacement space of the perovskite
structure. Subspaces are labelled by the corresponding Wyckoff symbol and site
symmetry group irrep. In cases where the
subspace is generated by a set of occupied Wyckoff positions and a vector
representation, the atom type is used as the label. For each subspace, the
symmetry labels at the high-symmetry $\vec k$-points are as in Ref.
\protect\onlinecite{cowley},
with conventions taken from
Ref. \protect\onlinecite{koster} and origin at A.}
\begin{tabular}{lcccc}
& $\Gamma$ & $R$ & $X$ & $M$ \\
\tableline
$A$ & $\Gamma_{15}$ & $R_{15}$ & $X'_2$,$X'_5$ & $M'_2$,$M'_5$ \\
$B$ & $\Gamma_{15}$ & $R'_{25}$ & $X_1$,$X_5$ & $M'_3$,$M'_5$  \\
$O_{x,x}$ & $\Gamma_{15}$ & $R'_{2}$,$R'_{12}$ & $X'_2$,$X_5$ &
$M_2$,$M'_3$,$M_4$ \\
$O_{x,y}$ & $\Gamma_{15}$,$\Gamma_{25}$ & $R_{25}$,$R_{15}$ &
$X_1$,$X_3$,$X_5$,$X'_5$ & $M_1$,$M_3$,$M_5$,$M'_5$\\
\end{tabular}
\label{tablea}
\end{table}
\narrowtext

\begin{table}
\caption{Band subspaces of the ionic displacement space of the $GeTe$ rocksalt
structure, as in Table I.
Symmetry label conventions are taken from Ref.
\protect\onlinecite{wigner}, with the origin at Te.}
\begin{tabular}{lcccc}
  & $\Gamma$ & $L$ & $X$ & $W$ \\
\tableline
 $Ge$ & $\Gamma_{15}$ & $L_1$,$L_3$ & $X'_4$,$X'_5$ & $W_1$,$W_3$ \\
 $Te$ & $\Gamma_{15}$ & $L'_2$,$L'_3$ & $X'_4$,$X'_5$ & $W'_2$,$W_3$ \\
\end{tabular}
\end{table}

\begin{table}
\caption{Band subspaces of the ionic displacement space of the $ZrO_2$ fluorite
structure, as in Table I. Symmetry label conventions are taken from Ref.
\protect\onlinecite{wigner}, with the origin at $Zr$.}
\begin{tabular}{lcccc}
  & $\Gamma$ & $L$ & $X$ & $W$ \\
\tableline
 $Zr$ & $\Gamma_{15}$ & $L'_2$,$L'_3$& $X'_4$,$X'_5$ & $W'_2$,$W_3$ \\
 $O$ & $\Gamma_{15}$,$\Gamma'_{25}$& $L_1$,$L_3$,$L'_2$,$L'_3$ &
$X_1$,$X_5$,$X'_2$,$X'_5$ & $W_1$,$W_3$,$W'_1$,$W_2$,$W'_2$ \\
$4b$,$\Gamma_{15}$ & $\Gamma_{15}$ & $L_1$,$L_3$& $X'_4$,$X'_5$ & $W_1$,$W_3$
\\
\end{tabular}
\end{table}

\mediumtext
\begin{table}
\caption{Band subspaces of the ionic displacement space of the A15 structure,
as in Table I. $A_{\parallel}$ denotes an A ion displacement along the chain
direction and $A_{\perp}$ denotes an A ion displacement perpendicular to the
chain direction.
For the $6b$ sites, the irreducible representations of the site symmetry group
are labelled by the component of a polar vector ($\vec u$) or an axial vector
$\vec R$ parallel to the chain ($\parallel$) or perpendicular and pointing
along the direction to the nearest B atom ($\perp,B$).  For the $6d$ sites, the
polar vector components are parallel or perpendicular to the ``chain''
directions defined by these sites. Symmetry labels at $\Gamma$ are taken from
Ref. \protect\onlinecite{wigner}~and at zone boundary points from
Ref. \protect\onlinecite{zoneb}, with the origin at B.}
\vskip 0.1in
\begin{tabular}{lcccc}
  & $\Gamma$ & $R$ & $X$ & $M$ \\
\tableline
 $6c,A_{\parallel}$ & $\Gamma_2$,$\Gamma_{12}$,$\Gamma_{15}$ & $4$&
$1$,$1$,$3$& $1$,$2$,$5$,$8$,$9$ \\
$6c,A_{\perp}$ & $\Gamma'_{15}$,$\Gamma_{15}$,$\Gamma'_{25}$,$\Gamma_{25}$&
$1$,$2$,$3$,$4$& $1$,$2$,$3$,$4$,$4$,$4$&
$2$,$3$,$6$,$7$,$9$,$9$,${10}$,${10}$\\
 $2a,B$ & $\Gamma_{15}$,$\Gamma_{25}$& $4$ & $1$,$3$,$4$&
$1$,$2$,$5$,$6$,${10}$\\
$6b,{\vec u_\parallel}$ & $\Gamma_{15}$,$\Gamma_{25}$& $4$& $1$,$3$,$3$&
$2$,$3$,$6$,$7$,$9$\\
$6b,{\vec u_{\perp,B}}$ & $\Gamma_{15}$,$\Gamma_{25}$& $1$,$2$,$3$&
$1$,$3$,$4$& $9$,$9$,$10$\\
$6b,{\vec R_\parallel}$ & $\Gamma'_{15}$,$\Gamma'_{25}$& $4$& $2$,$3$,$3$&
$2$,$3$,$6$,$7$,$10$\\
$6b,{\vec R_{\perp,B}}$ & $\Gamma'_{15}$,$\Gamma'_{25}$& $1$,$2$,$3$&
$2$,$3$,$4$& $9$,$10$,$10$\\
$6d,{\vec u_\parallel}$ & $\Gamma_2$,$\Gamma_{12}$,$\Gamma_{15}$ & $4$&
$1$,$1$,$3$& $1$,$2$,$5$,$8$,$9$ \\
$6d,{\vec u_\perp}$ &
$\Gamma'_{15}$,$\Gamma_{15}$,$\Gamma'_{25}$,$\Gamma_{25}$& $1$,$2$,$3$,$4$&
$1$,$2$,$3$,$4$,$4$,$4$& $2$,$3$,$6$,$7$,$9$,$9$,${10}$,${10}$
\end{tabular}
\end{table}

\begin{figure}
\caption{Eigenvalues are shown as a function of $k$ for the two branches of
normal modes of the diatomic chain with
${\gamma\over\alpha}=0.25$.
At $k=0$ and $k=\pi$, each branch is labelled by the appropriate irreducible
representation of the space group and the corresponding displacement pattern is
shown in the adjacent sketch. A ions are shown
as open circles and B ions as solid circles. \label{abmodes_fig}}
\end{figure}

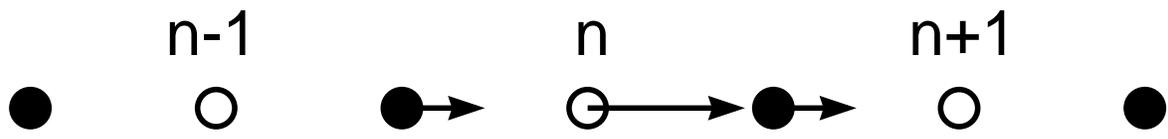
\begin{figure}
\caption{The displacement pattern of the Wannier function $w_{\Lambda n}$,
corresponding to the lower branch in Fig.
\protect\ref{abmodes_fig}, is shown by the arrows. The pattern is odd around
the central A ion. The arrows for the outermost A and B ions are smaller than
the ion symbols. \label{aodd_fig}}
\end{figure}

\begin{figure}
\caption{Dispersion ${\omega^2(k)\over S(k)}$ for the lower branch computed
using the approximate Wannier function fit to $k=0$ (lower solid curve)
compared with the true dispersion (dashed curve). The upper solid curve shows
the overlap matrix element $S$ as a function of $k$.
For comparison, the upper dashed line is the unit normalization $S=1$.
\label{abwapp_fig}}
\end{figure}

\begin{figure}
\caption{Displacement patterns for the innermost coordination shells
of representative Wannier basis vectors in each of the four band subspaces of
the perovskite structure. The open squares represent the A ions, the shaded
squares the B ions, and the open and shaded circles represent inequivalent
oxygen ions.
The four lines in the figure correspond to the subspaces A, B, $O_{x,x}$ and
$O_{x,y}$. In the first structure in each line, the displacement of each set of
inequivalent ions can be specified independently. \label{pattern_fig}}
\end{figure}

\begin{references}
\bibitem{isotropy} The tedium of generating these tables for a complicated
crystal structure can be relieved through the use of the
computer program ISOTROPY, described in H. T. Stokes, Ferroelectrics (in
press).
\end{references}
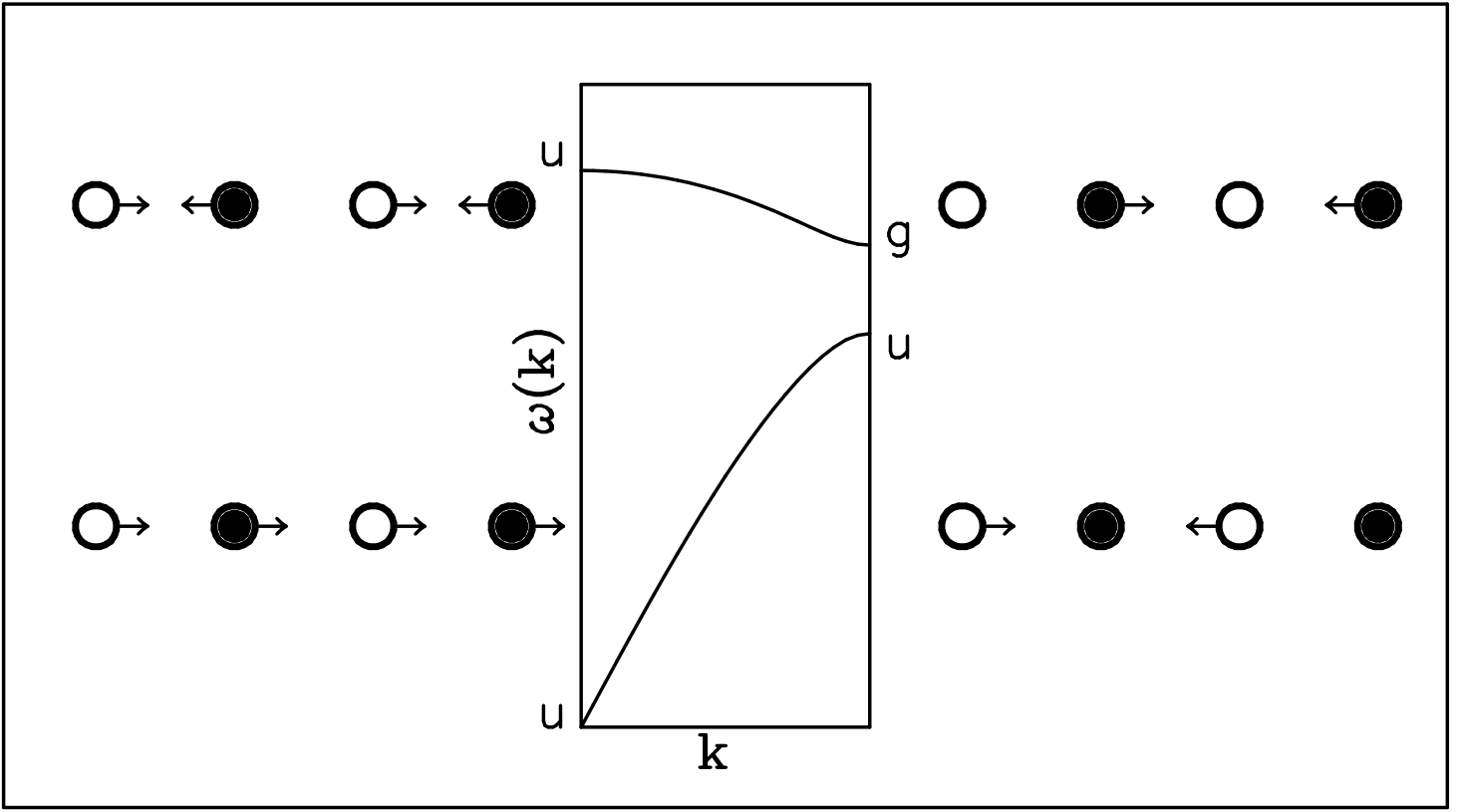
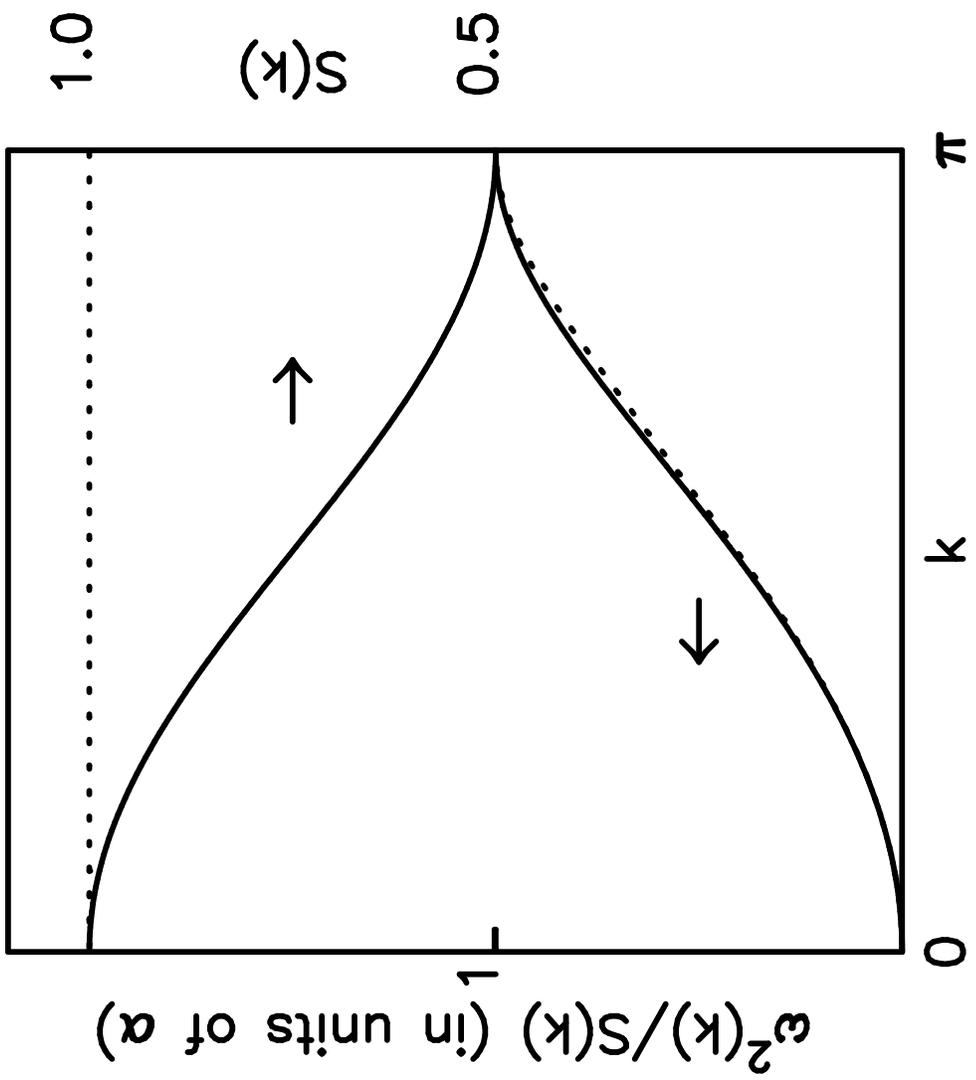
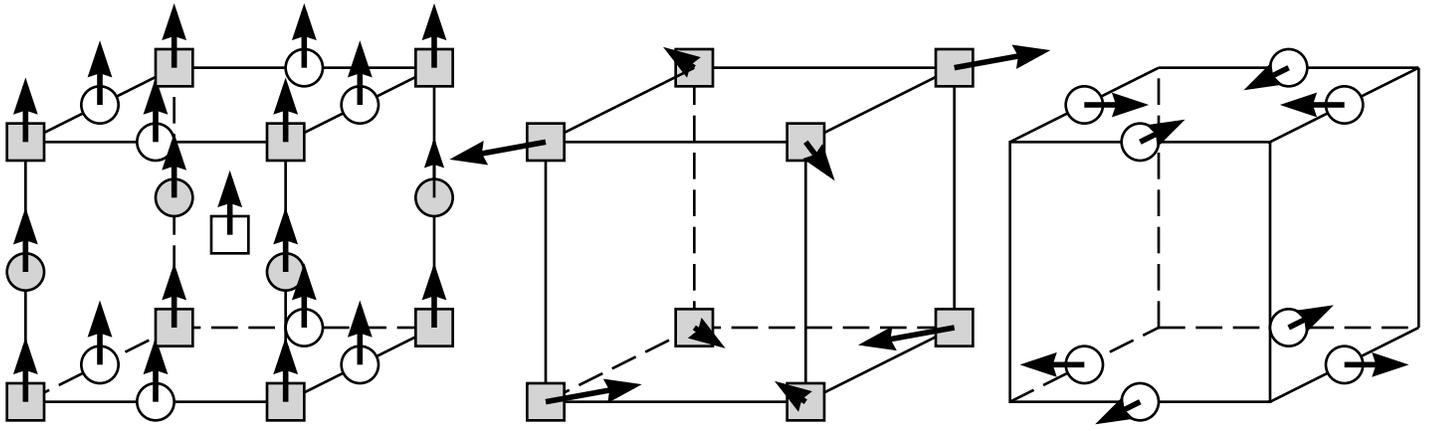
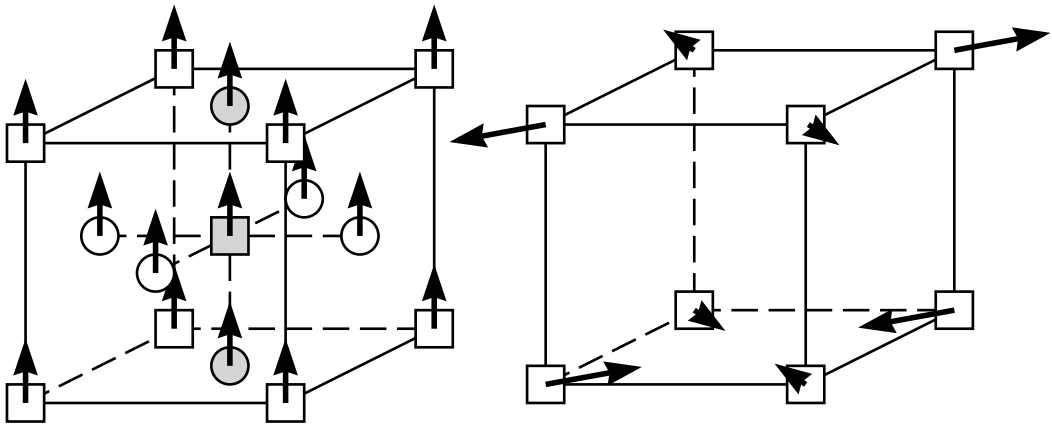
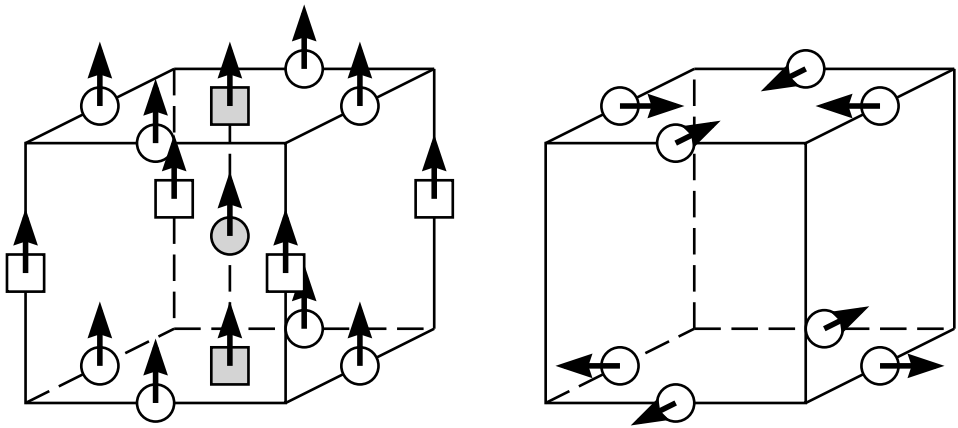
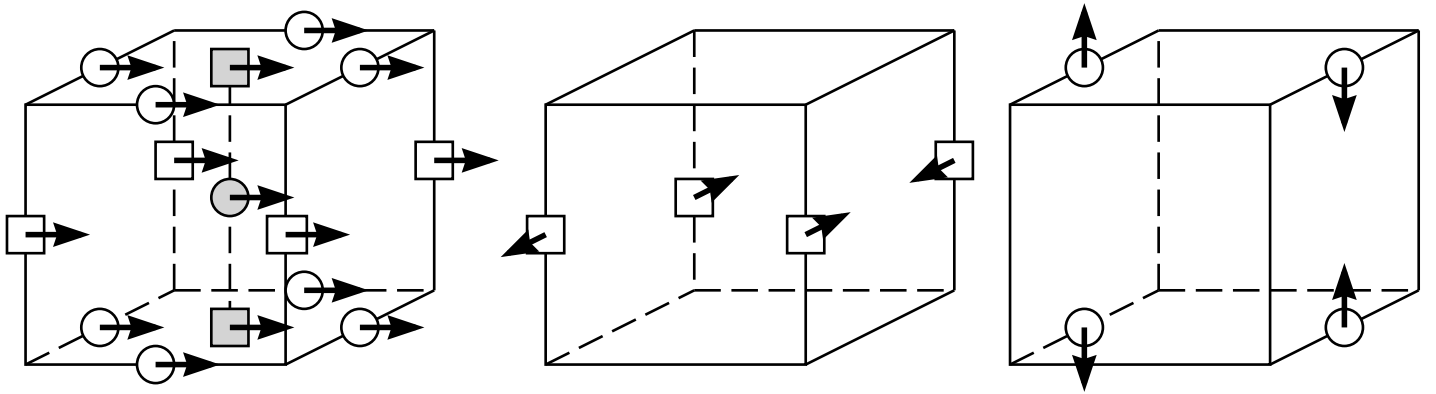
\end{document}